\documentclass[graybox]{svmult}

\usepackage{mathptmx}       
\usepackage{helvet}         
\usepackage{courier}        
\usepackage{type1cm}        
\usepackage{makeidx}         
\usepackage{graphicx}        
\usepackage{multicol}        
\usepackage[bottom]{footmisc}

\makeindex             

\usepackage{natbib}

\usepackage{amssymb,amsmath}

\begin{document}

\title*{Synergies between Asteroseismology and Exoplanetary Science}
\author{Daniel Huber}
\institute{Daniel Huber \at Institute for Astronomy, University of Hawaii, 2680 Woodlawn Drive, Honolulu, HI 96822, USA,\\
\email{huberd@hawaii.edu}\\ \\
Sydney Institute for Astronomy (SIfA), School of Physics, University of Sydney, NSW 2006, Australia\\ \\
Stellar Astrophysics Centre (SAC), Department of Physics and Astronomy, Aarhus University, Ny Munkegade 120, DK-8000 Aarhus C, Denmark}
%
%
\maketitle

\abstract{Over the past decade asteroseismology has become a powerful method to systematically characterize host stars and dynamical architectures of exoplanet systems. In this contribution I review current key synergies between asteroseismology and exoplanetary science such as the precise determination of planet radii and ages, the measurement of orbital eccentricities, stellar obliquities and their impact on hot Jupiter formation theories, and the importance of asteroseismology on spectroscopic analyses of exoplanet hosts. I also give an outlook on future synergies such as the characterization of sub-Neptune-size planets orbiting solar-type stars, the study of planet populations orbiting evolved stars, and the determination of ages of intermediate-mass stars hosting directly imaged planets.}

\section{Introduction: Know the star, know the planet}

Exoplanetary science has undergone a revolution over the past two decades, driven by ground-based Doppler surveys and high-precision, space-based photometry from missions such as \textit{CoRoT} \citep{baglin09} and \textit{Kepler} \citep{borucki10}. At the time of writing nearly 3500 confirmed exoplanets are known, and future space-based missions such as \textit{TESS} \citep{ricker14} in combination with ground-based efforts are expected to continue this revolution over the coming decades.

The wealth of exoplanet discoveries has uncovered several important questions: How did gas-giant planets in close-in orbits (hot Jupiters) form? What are the origin and compositions of sub-Neptune-size planets, for which we have no equivalent in the solar system? What are the occurrence rates of exoplanets as a function of their size, mass, orbital architecture, as well as their host star spectral type and evolutionary state? Do habitable planets exist outside our solar system?

Our ability to answer these questions depends strongly on our understanding of the host stars. This is primarily due to the fact that the majority of exoplanet detections are indirect --- more than 98\% of all exoplanets known to date were discovered using transits, the Doppler method, or microlensing, all of which measure properties of planets relative to the host star. Thus, constraining the physical properties of planets is often limited by the characterization of stars. \textit{Indeed, for 99\% of all planet candidates detected by Kepler the uncertainty in the planet radius is currently dominated by the uncertainty in the radius of the host star.} In addition to placing planet properties on an absolute scale, host star characteristics are also crucial to understand the planetary environments such as the extent of the habitable zone \citep{kane14}.

\begin{figure}[t]
\begin{center}
\resizebox{\hsize}{!}{\includegraphics{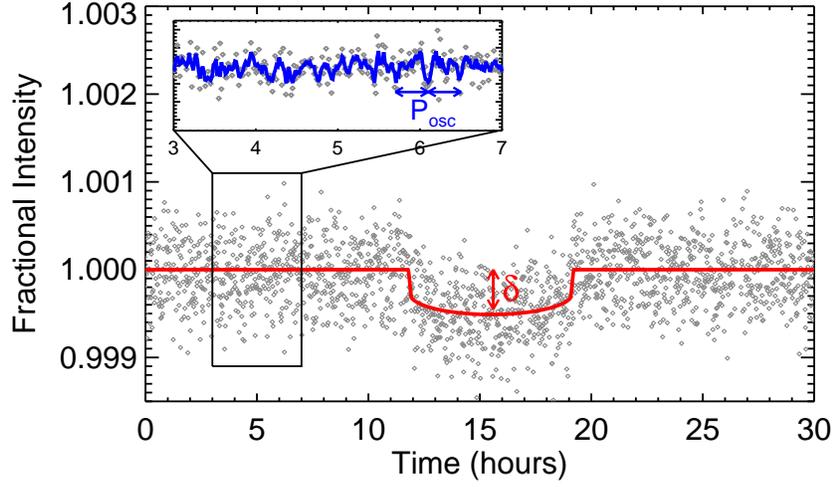}}
 \caption{\textit{Kepler} short-cadence light curve showing a single transit of Kepler-36c.  The red solid line is the transit model from \citet{carter12}, and the inset shows the oscillations of the host star. The transit depth $\delta$ yields the size of the planet relative to the star, and the oscillation periods ($P_{\rm osc}$) can be used to independently measure the size of the star.}
\label{fig:kepler36}
\end{center}
\end{figure}

The requirement for continuous high-precision monitoring has enabled a fortuitous synergy between asteroseismology and exoplanetary science, since the data can be simultaneously used to detect exoplanets and study stellar oscillations (see Fig.~\ref{fig:kepler36}). In this review I will discuss some key synergies between both fields, and conclude with an outlook of what future synergies we can expect from current and future ground- and space-based facilities such as SONG, K2, \textit{TESS}, \textit{PLATO} and \textit{WFIRST}.

\section{Characterization of exoplanets}

\subsection{The connection between transits and mean stellar density}

The primary observable for exoplanet transits is the transit depth\footnote{The notation $\Delta F$ will be used hereafter to denote transit depth.}, $\Delta F$, which for the simplified case of a uniformly bright stellar disk is related to the size of the planet ($R_{\rm P}$) and the size of the star ($R_{\star}$) as:
\begin{equation}
\Delta F = \left(\frac{R_{\rm P}}{R_{\star}}\right)^{2} \, .
\end{equation}

\begin{figure}[t]
\begin{center}
\resizebox{9cm}{!}{\includegraphics{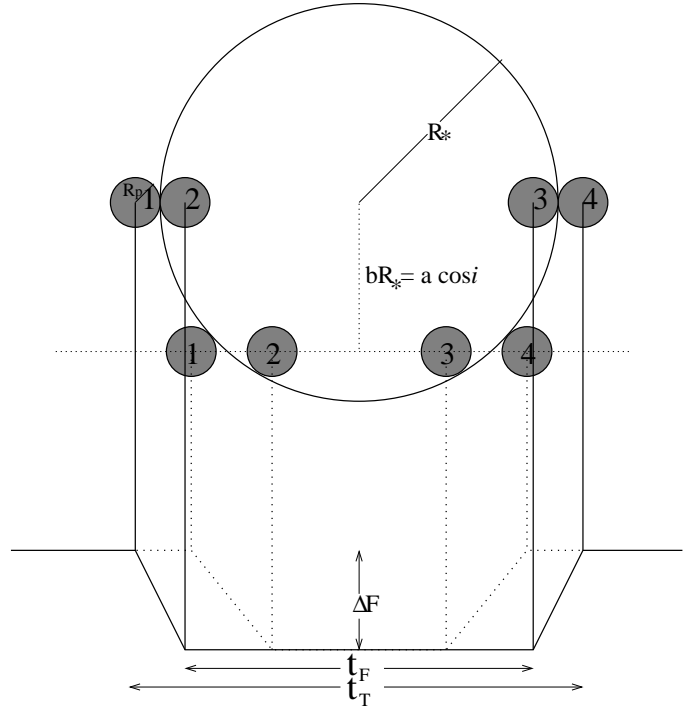}}
 \caption{Schematic transit light curves (solid and dotted lines on the bottom) and corresponding star-planet geometry (top). The four transit contact points are shown for both transits. The transit depth, $\Delta F$, total transit duration, $t_T$, and transit duration between ingress and egress, $t_F$, are shown for the solid transit light curve. Also defined is the impact parameter, $b$. From \citet{seager03}.}
\label{fig:transit}
\end{center}
\end{figure}

Accurate measurements of $R_{\rm P}/R_{\star}$, however, are typically complicated by degeneracies between the transit depth, transit duration, impact parameter, limb darkening, and the size of the star (see Fig.~\ref{fig:transit}). For example, for fixed $R_{\rm P}/R_{\star}$ a larger impact parameter will lead to a shallower transit (due to limb darkening) with shorter duration. The same transit duration and depth, however, could likewise be caused by a smaller planet orbiting a smaller star with a lower impact parameter.

This degeneracy can be broken with independent knowledge of the host star density. Assuming $R_{\rm P} \ll R_{\star} \ll a$ (where $a$ is the orbit semi-major axis) and circular orbits, it can be shown that \citep{winn10}:
\begin{equation}
\frac{a}{R_{\star}} = \frac{2 \, \Delta F^{1/4}}{\pi} \frac{P}{\sqrt{t_T^2 - t_F^2}} \, .
\end{equation}
Here, $P$ is the orbital period, $t_T$ is the total transit duration and $t_F$ is the transit duration between ingress and egress, as illustrated in Fig.~\ref{fig:transit}. Using Kepler's third law, $a/R_{\star}$ can be expressed as
\begin{equation}
\frac{a}{R_{\star}} = \left(\frac{P^{2} \, G}{4 \pi^{2}} \frac{M_{\star}}{R_{\star}^{3}}\right)^{1/3} \, ,
\end{equation}
and hence:
\begin{equation}
\rho_{\star} = \frac{3\pi}{G \, P^2} \left(\frac{a}{R_{\star}}\right)^{3} \, .
\label{equ:dur}
\end{equation}
The mean stellar density is therefore directly related to quantities which can be  measured from a transit light curve \citep{seager03}. 

Equation (\ref{equ:dur}) is of key importance for the synergy between asteroseismology and exoplanetary science. Since asteroseismology measures the mean stellar density with a typical precision of a few percent or less, the combination of stellar oscillations and transits can be used to remove degeneracies when fitting exoplanet transits and accurately measure transit parameters. This is particularly important for small planets with low-$SNR$ transits, for which ingress and egress durations often cannot be accurately measured and hence constraining $a/R_{\star}$ independently of $\rho_{\star}$ is difficult.

\subsection{The importance of precise exoplanet radii}

Precise host star radius measurements are important for understanding the composition of planets. Composition models depend sensitively on radius, especially in the regime of sub-Neptune-size planets, and density measurements from transit and Doppler surveys have indicated a threshold between mostly rocky and gaseous planet compositions of $\approx$\,1.6\,$R_{\oplus}$ \citep{weiss14,rogers15}. Uncertainties in planet radii due to indirect stellar characterization methods, however, have often led to ambiguities when interpreting exoplanet detections. For example, the $\approx$\,20\% radius uncertainty for Kepler-452b prevented firm conclusions about whether the planet, which orbits a G-type host star within the habitable zone, is indeed rocky \citep{jenkins15}. 

\begin{figure}[t]
\begin{center}
\resizebox{\hsize}{!}{\includegraphics{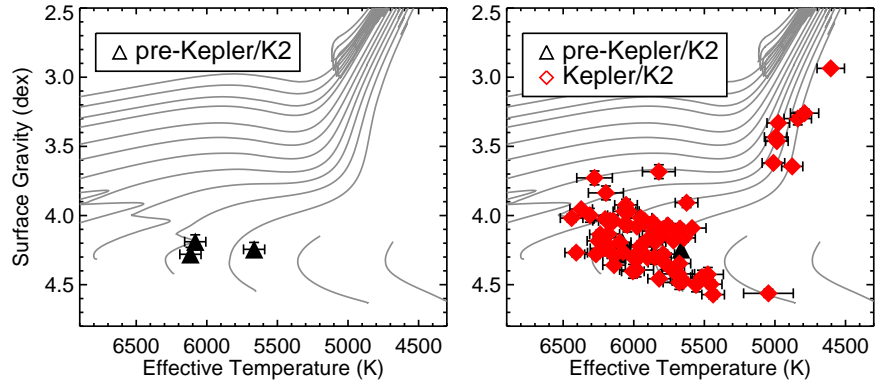}}
\caption{Surface gravity versus effective temperature for exoplanet-host stars with asteroseismic detections before (left panel) and after (right panel) the launch of \textit{Kepler}/K2. Gray lines show solar-metallicity evolutionary tracks from the \textsc{basti} database \citep{basti}.}
\label{hrd}
\end{center}
\end{figure}

Asteroseismology has provided some of the most precise characterizations of exoplanets to date. The first asteroseismic studies of exoplanet-host stars were performed using ground-based, radial-velocity observations of $\mu$\,Ara \citep{bazot05,bouchy05}, space-based photometry using the \textit{Hubble Space Telescope} of HD~17156 \citep{gilliland11} and \textit{CoRoT} photometry of HD~52265 \citep{ballot11b,lebreton14}. The launch of \textit{Kepler} led to a revolution in the synergy between asteroseismology and exoplanetary science, with over 70 confirmed \textit{Kepler} exoplanet-host stars (see Fig.~\ref{hrd}). This large sample allowed the first systematic precise characterization of planets in the \textit{Kepler} sample \citep{huber13}, including planet radii measured to $\approx$\,1\,\% \citep{ballard14}, as well as investigations of the effects of stellar incident flux on the radius distribution of close-in planets \citep{lundkvist16}.

More recent studies have focused not only on measuring global asteroseismic quantities 
(which are sensitive to densities, masses and radii) but also systematic 
modeling of individual oscillation frequencies, which allows precise constraints on  
stellar ages \citep{silva15,davies15}. One of the most remarkable discoveries so far is Kepler-444, which consists of a K dwarf of age $11.2\pm1.0$\,Gyr hosting
five sub-Earth-size planets with orbital periods of less than 10 days \citep{campante15}. Kepler-444 demonstrated that sub-Earth-size planets have existed for most of the history of our Universe, and the discovery of a pair of low-mass companions in a highly eccentric orbit furthermore showed that the formation of small planets appears to be robust against early truncation of the protoplanetary disk \citep{dupuy16}.

\section{Orbital eccentricities of exoplanets}

Orbital eccentricities play a key role in many areas of exoplanetary science, ranging from studies of the dynamics of multiplanet systems to the determination of the fraction of time a planet spends within the habitable zone. Traditionally, eccentricities can be measured through Doppler velocities, secondary transits, or transit-timing variations. However, these methods are either only applicable for relatively large gas-giant planets, or a small subset of multiplanet systems for which effects of eccentricity and mass can be successfully disentangled \citep[e.g.,][]{lithwick12,hadden14}.

\begin{figure}[t]
\begin{center}
\resizebox{10cm}{!}{\includegraphics{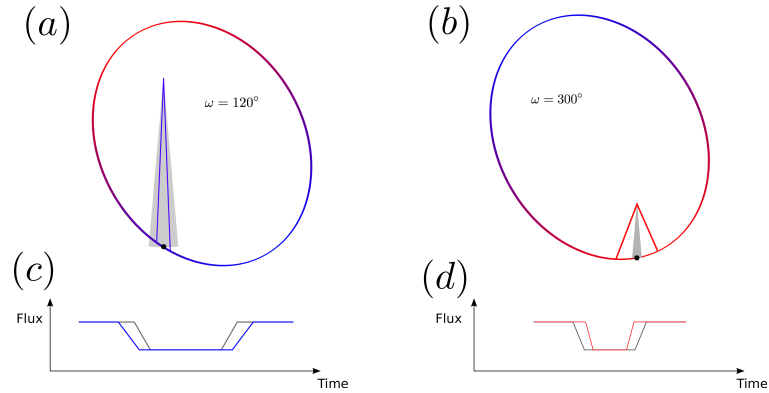}}
\caption{Planetary orbit of eccentricity 0.6 with two different angles of periastron (top panels) and the corresponding observed transits (bottom panels). Red and blue colors correspond to the fast and slow part of the orbit, respectively. Observed transit durations are longer (left panels) or shorter (right panels) compared to circular orbits (gray) depending on the eccentricity and argument of periastron of the orbit. From \citet{vaneylen15}.}
\label{orbit}
\end{center}
\end{figure}

The combination of transit photometry and asteroseismology has opened up a powerful method to systematically measure orbital eccentricities of transiting planets. Since the eccentricity and orientation of the orbit to the observer control the transit duration (see Fig.~\ref{orbit}), the ratio of the mean stellar density assuming a circular orbit (Eq~\ref{equ:dur}) and true mean stellar density are related as \citep[e.g.,][]{kipping10}:

\begin{equation}
\frac{\rho_{\star}}{\rho_{\star, \rm{transit}}} = \frac{(1-e^{2})^{3/2}}{(1+e \sin{\omega})^{3}} \, .
\label{equ:ecc}
\end{equation}
Here, $e$ is the eccentricity and $\omega$ is the argument of periastron. Equations (\ref{equ:dur}) and (\ref{equ:ecc}) demonstrate that if an independent measurement of $\rho_{\star}$ is available (for example, from asteroseismology), transits can be used to directly constrain the eccentricity of a planet without radial-velocity observations. Importantly, an accurate measurement of $\rho_{\star, \rm{transit}}$ requires an accurate estimate of the ingress and egress times.

\begin{figure}[t]
\begin{center}
\resizebox{\hsize}{!}{\includegraphics{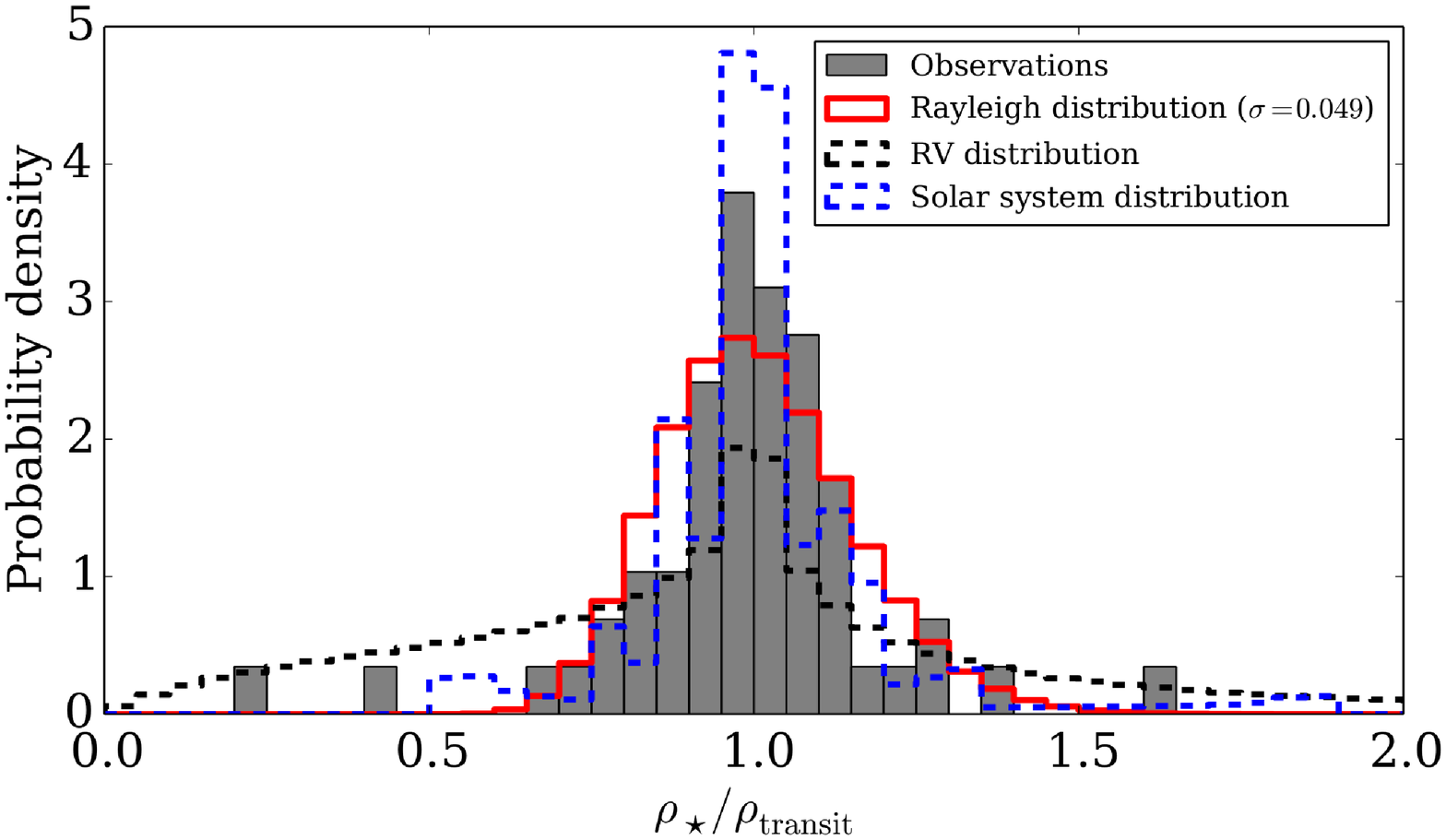}}
\caption{Ratio of the asteroseismic mean stellar density and the density measured from transits assuming a circular orbit for 28 \textit{Kepler} multiplanet host stars (gray). Corresponding distributions of solar system planets and planets detected by radial-velocity 
surveys are shown as blue and black dashed histograms, respectively. 
From \citet{vaneylen15}.}
\label{fig4}
\end{center}
\end{figure}

The first systematic study of eccentricities using asteroseismic densities concentrated on the identification of false positives in the \textit{Kepler} planet candidate sample by comparing $\rho_{\star}$ and $\rho_{\star, \rm{transit}}$, yielding a significantly higher false-positive rate for red-giant-host stars \citep{sliski14}. A subsequent study by \citet{vaneylen15} focused on 28 multiplanet systems, which are expected to have a small false-positive rate \citep{lissauer12}. Figure \ref{fig4} shows a histogram of the derived ratios between transit and seismic density (left hand side of Eq.~\ref{equ:ecc}) for their sample compared to the solar system and a sample of planets with eccentricities from radial-velocity surveys. The asteroseismic sample (red solid line) is consistent with circular orbits, similar to the solar system (blue dashed histogram), but in stark contrast to the radial-velocity sample 
(black dashed line). Since \textit{Kepler} multiplanet systems include mostly small, 
low-mass planets compared to the more massive planets probed by Doppler surveys, this indicates that low-mass planets are preferentially on 
circular orbits. This conclusion is of great importance since circular orbits are frequently assumed when modeling exoplanets in the habitable 
zone \citep{barclay13,borucki13,quintana14,jenkins15} or when estimating the detection completeness for planet occurrence studies \citep[e.g.,][]{howard11,dong13b,petigura13b,burke15}. 

Expanding such studies holds promise to further constrain the dynamics of exoplanet systems using asteroseismology. For example, \citet{xie16} recently used stellar densities derived from spectroscopy to show that while multiplanet systems are indeed preferentially circular, single systems appear to show significantly higher eccentricities even for small (sub-Neptune-size) planets. Asteroseismic studies of systems with single planets would be valuable to independently confirm this result with a smaller, but higher precision sample. 

\begin{figure}[t]
\begin{center}
\resizebox{10cm}{!}{\includegraphics{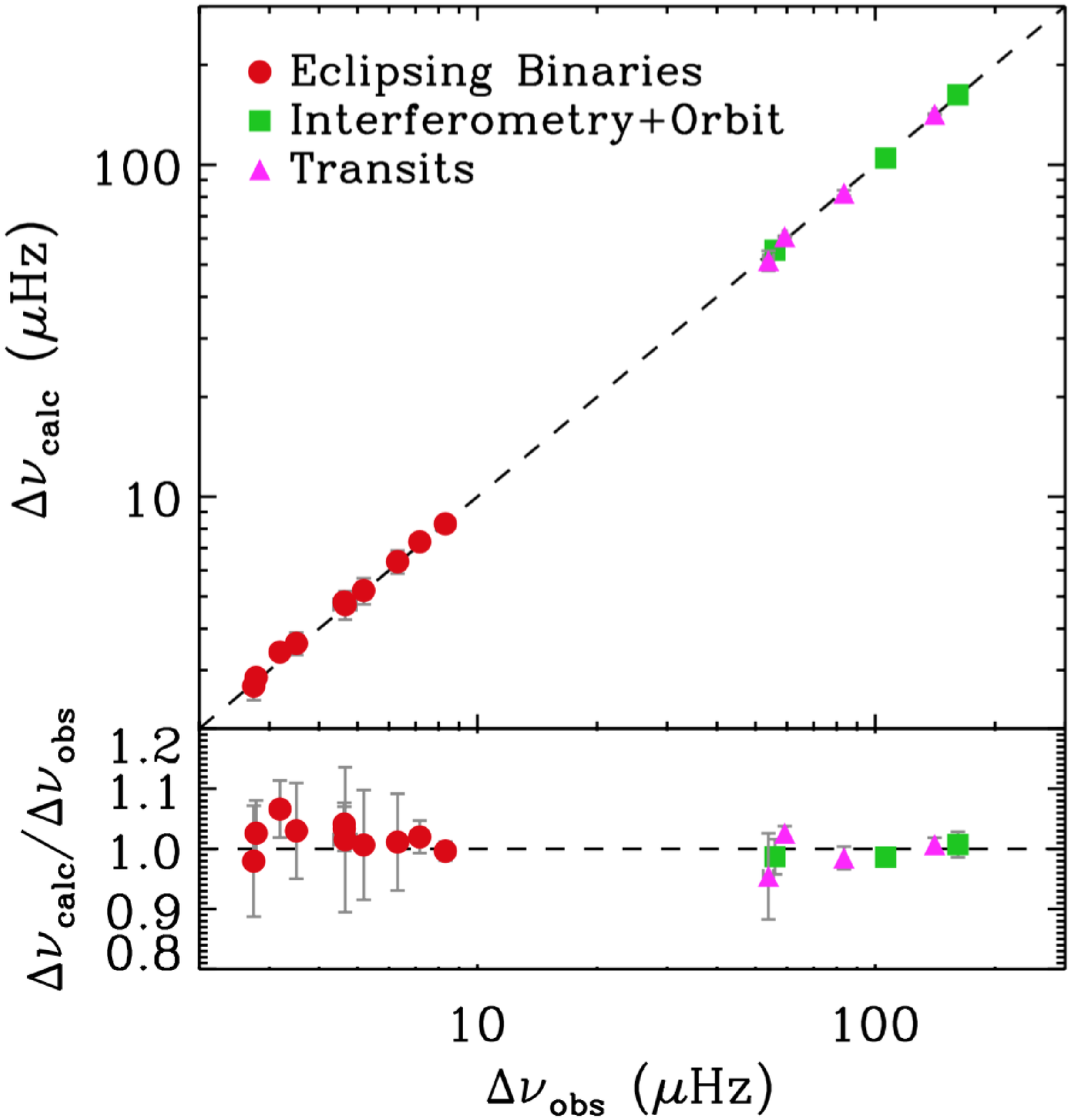}}
\caption{Comparison of the mean stellar density from independent methods and as calculated from the asteroseismic scaling relation for the large frequency separation ($\Delta\nu$). Transit-derived densities account for more than half of the comparison values for subgiant and main-sequence stars.}
\label{dnuscaling}
\end{center}
\end{figure}

Transiting exoplanets for which the eccentricity can be measured independently (for example through radial velocities) can also be used as an independent test of asteroseismic densities calculated from the scaling relation for the large frequency separation ($\Delta\nu$). This is particularly valuable since the $\Delta\nu$ scaling relation has found widespread use for calculating stellar properties for thousands of stars in the era of ``ensemble asteroseismology'' \citep{kallinger10,chaplin13}. Figure \ref{dnuscaling} shows a comparison of the mean stellar density calculated from the $\Delta\nu$ scaling relation and from dynamically measured densities from double-lined eclipsing binaries \citep{frandsen13,gaulme16}, interferometric orbits \citep[Procyon and $\alpha$\,Cen\,A+B; see][and references therein]{bruntt10}, as well as transiting exoplanets with known eccentricities: HD~17156 \citep{gilliland11,nutzman11}, TrES-2 \citep{southworth11,barclay12}, HAT-P-7 \citep{cd10,southworth11}, and Kepler-14 \citep{southworth12,huber13}. Transit-derived densities account for more than half of the currently available comparison values for subgiant and main-sequence stars, and empirically demonstrate that the $\Delta\nu$ scaling relation is accurate to about $\approx$\,3\,\% \citep[see also][]{huber14b}.

\section{Obliquities of exoplanet systems}
\label{sec:obl}

\begin{figure}[t]
\begin{center}
\resizebox{6cm}{!}{\includegraphics{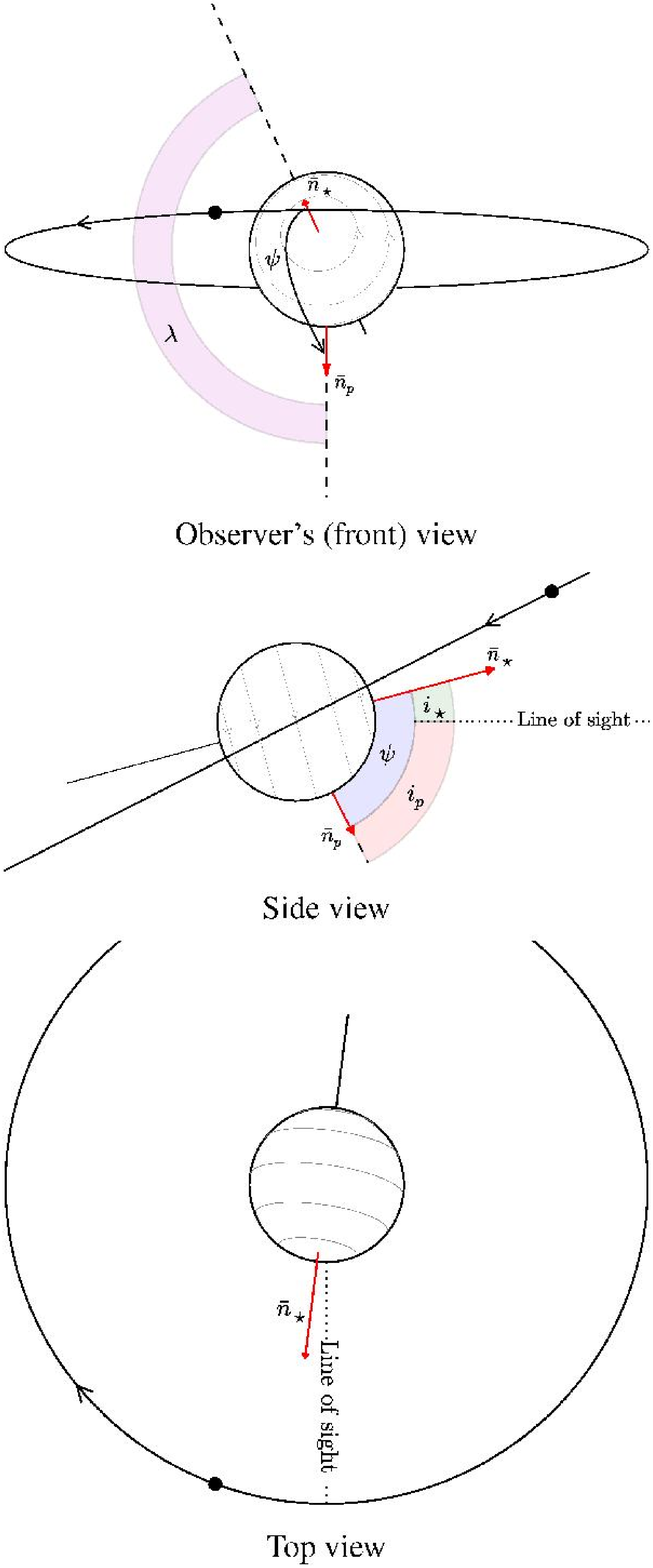}}
\caption{Graphical illustration of the obliquity $\psi$, the sky-projected spin-orbit angle $\lambda$, the line-of-sight stellar inclination $i_{\star}$, and the line-of-sight orbit inclination $i_{\rm p}$. Note that the top panel shows $\lambda=155^{\circ}$, while the middle panel shows $\lambda=180^{\circ}$. From \citet{lund14}.}
\label{obliquity}
\end{center}
\end{figure}

The obliquity $\psi$ is the angle between the host star rotation axis and the planetary orbital axis, and can be calculated as \citep{fabrycky09}:
\begin{equation}
\cos \psi = \sin i_{\star} \cos \lambda \sin i_{\rm p} +  \cos i_{\star} \cos i_{\rm p} \, .
\end{equation}
Here, $\lambda$ is the sky-projected spin-orbit angle, $i_{\rm p}$ is the angle between the line of sight and the orbital axis of the planet, and $i_{\star}$ is the inclination of the rotation axis to the line of sight of the observer. Figure \ref{obliquity} shows a graphical illustration of these angles for the HAT-P-7 system following \citet{lund14}.

For transiting exoplanets, $i_{\rm p}$ can be typically constrained from the transit light curve, while $\lambda$ can be measured through spectroscopic in-transit observations (the Rossiter--McLaughlin effect). Asteroseismic observations of the relative heights of rotationally split multiplets can be used to provide the measurement of the line-of-sight inclination of the stellar rotation $i_{\star}$ \citep{gizon03}. Thus, the combination of transits, Doppler velocities, and asteroseismology allow to uniquely measure the obliquity of exoplanet systems \citep{benomar14,lund14}. Importantly, the measurement of $i_{\star}$ using asteroseismology is independent of planet size, and hence can be used to constrain the obliquity even for systems with small planets in which Rossiter--McLaughlin measurements are typically not feasible. For transiting planets, a low stellar inclination in most cases automatically yields a misalignment of the orbital plane and the stellar equatorial plane (a high obliquity), while a value of near $90^{\circ}$ for $i_{\star}$ implies that the star and the planets are likely (but not necessarily) aligned.

\begin{figure}[t]
\begin{center}
\resizebox{\hsize}{!}{\includegraphics{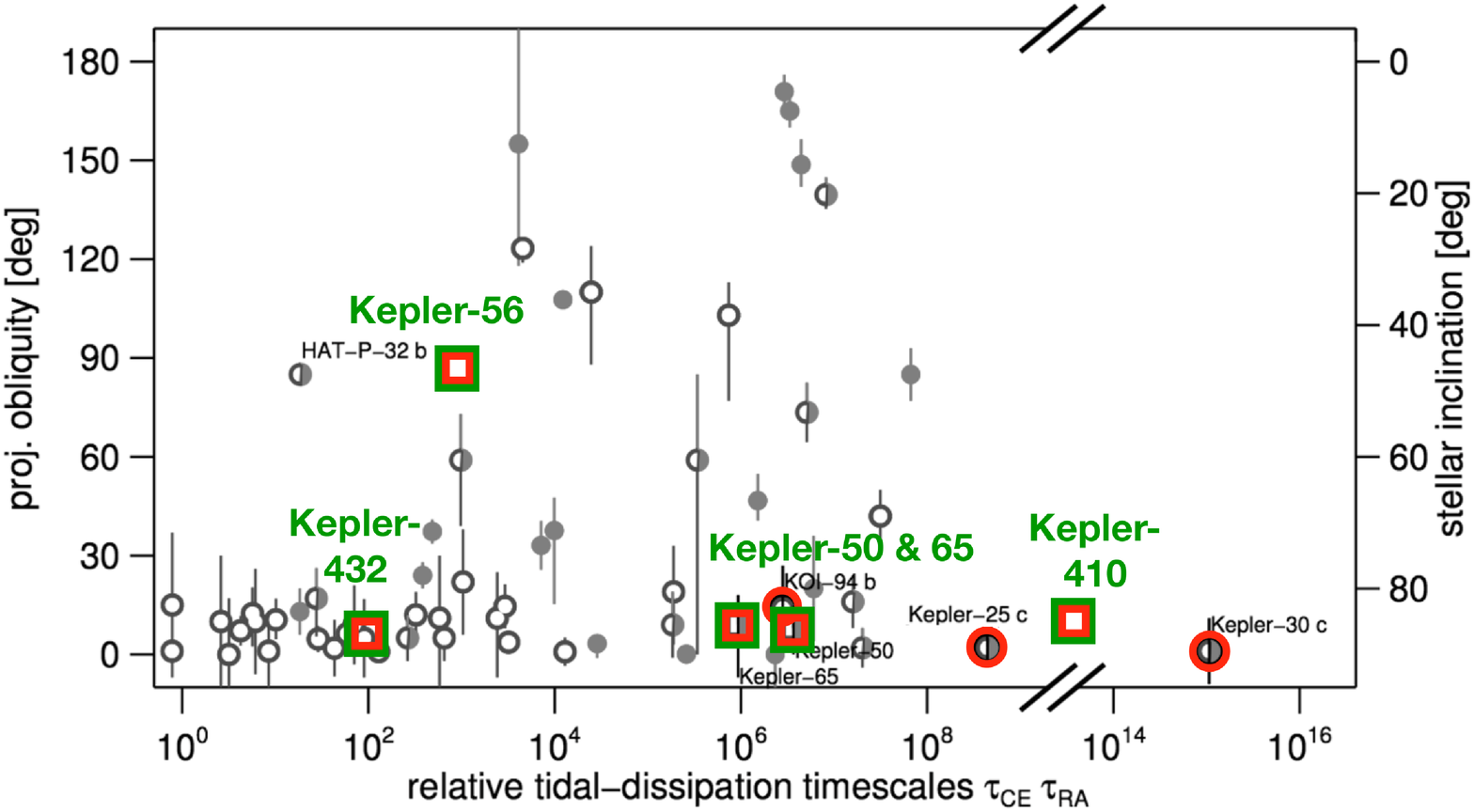}}
\caption{Projected obliquity (left ordinate) and stellar inclination (right ordinate) 
versus relative tidal 
dissipation timescale for exoplanet systems. Systems with short dissipation timescales 
are expected to have been realigned even if they were misaligned by the formation 
process, while systems with long dissipation timescales are expected to preserve their 
configuration. Multiplanet systems without hot Jupiters are highlighted 
by red circles, and 
systems with inclinations measured using asteroseismology are highlighted with green 
squares. Positions for Kepler-56, Kepler-410 and Kepler-432 are 
approximate only. Adapted from \citet{albrecht13}.}
\label{obliquity2}
\end{center}
\end{figure}

Obliquities have played a key role in constraining the formation mechanism for hot Jupiters, one of the longest standing problems in exoplanetary science. Hot Jupiters are typically thought to form at large orbital distances beyond the snow line, and subsequently migrate to the close-in orbits where they are currently observed \citep[although in-situ formation has also been suggested; see][]{batygin16}. Two possible mechanisms have been proposed: migration of the planet through the protoplanetary disk \citep{lin96} or dynamical perturbations such as planet-planet scattering \citep{chatterjee08} or Kozai--Lidov oscillations \citep{fabrycky07} which cause the planet to attain a high orbital eccentricity, followed by shrinking and circularization of the orbit through tidal interactions (often referred to as high-eccentricity migration). 

The observation that hot Jupiters show a wide range of obliquities \citep{johnson09,winn10} has been interpreted as evidence for a dynamically violent formation scenario, thus favoring high-eccentricity migration as the dominant formation mechanism. However, this conclusion relies on the assumption that the stellar equator and the protoplanetary disk are initially aligned, and thus that the high obliquity observed today is indeed a consequence of dynamical interactions during the migration process. Key tests for this assumption are multiplanet systems which, if primordial alignments are common, should predominantly show low obliquities. 

Asteroseismology has played an important role for testing this assumption since seismic inclination measurements are independent of planet size, and hence can be applied to multiplanet systems with small planets. Figure \ref{obliquity2} shows the projected obliquity or stellar inclination for exoplanet systems as a function of relative tidal dissipation timescale, which is a proxy for how quickly a system can be realigned by tidal interactions if it was initially misaligned by the formation process \citep{albrecht12}. In line with expectations from high-eccentricity migration, hot-Jupiter systems with intermediate dissipation timescales are frequently observed to have high obliquities, while coplanar multiplanet systems without hot Jupiters have mostly low obliquities \citep[e.g.,][]{sanchis13} despite long tidal dissipation timescales. Over half of the constraints for multiplanet systems come from asteroseismology \citep{chaplin13c,vaneylen14,quinn15}.

Asteroseismology has also yielded the first intriguing counterexample for the observed trend of well-aligned multiplanet systems. Kepler-56, a red giant hosting two transiting planets confirmed through transit-timing variations \citep{steffen12}, revealed an inclination of $i_{\star}=47^{\circ}\pm6^{\circ}$, demonstrating the first  spin-orbit misalignment in a multiplanet system \citep{huber13b}. Subsequent follow-up studies have confirmed that the misalignment is likely caused by the torque of a third planet on a wide orbit \citep{li14,otor16,gratia17}, and that such a configuration could be consistent with a primordial misalignment \citep{matsakos17}. Future asteroseismic inclination measurements will be needed to determine whether  
spin-orbit misalignments in multiplanet systems are common, and whether high obliquities are indeed tracers of dynamical formation history of hot Jupiters.

\section{Chemical abundances of exoplanet-host stars}

Chemical abundances of exoplanet-host stars are tracers of the primordial composition of protoplanetary disks, and hence provide valuable clues about which conditions favor planet formation. For example, it is well established that gas-giant planets predominantly form around metal-rich stars \citep{gonzalez97,fischer05}, whereas small planets form independently of host star metallicity \citep{buchhave12}. Going beyond metallicities, intriguing abundance differences in volatile and refractory elements between the Sun and solar twins with and without planets have been observed \citep{melendez09,ramirez09}, although the link of these patterns to terrestrial planet formation is still being debated \citep[e.g.,][]{adibekyan14}.

\begin{figure}[t]
\begin{center}
\resizebox{11cm}{!}{\includegraphics{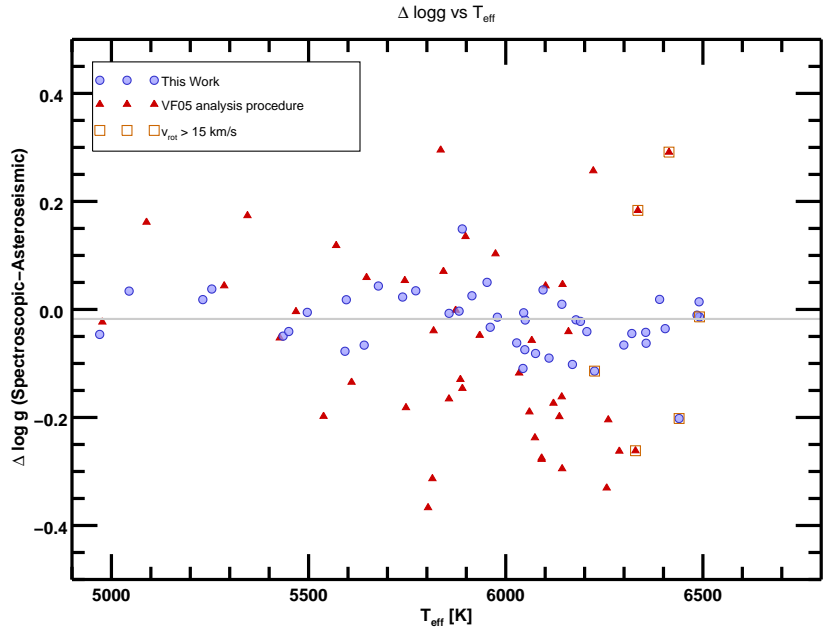}}
\caption{Difference between asteroseismic and spectroscopic surface gravities before (red) and after (blue) improving the spectroscopic modeling procedure based on asteroseismic constraints on surface gravity. From \citet{brewer15}.}
\label{fig:brewer}
\end{center}
\end{figure}

Asteroseismology does not directly probe atmospheric abundances, but the combination of asteroseismology and spectroscopy can significantly improve our understanding of host star compositions. This is mainly due to the fact that bulk atmospheric parameters ($T_{\rm eff}$, $\log g$, [Fe/H], microturbulence) are often heavily correlated, which can lead to systematic errors in particular for spectral synthesis methods \citep{torres12,huber13}, while spectral line analysis methods are typically less affected \citep{mortier14}. Recent efforts have shown that using $\log g$ from asteroseismology to inform spectroscopic modeling methods can significantly increase the accuracy of spectroscopic surface gravities without external constraints (see Fig.~\ref{fig:brewer}), thus also leading to more accurate abundances. Using stars with asteroseismology (ideally in combination with interferometry, which also yields an external constraint on $T_{\rm eff}$) as spectroscopic benchmarks promises to extend high-precision abundance work from solar twins to stars in different evolutionary states.

\section{Future prospects}

The asteroseismology revolution initiated by \textit{CoRoT} and \textit{Kepler} is set to continue over the coming decades with the launch of \textit{TESS} \citep{ricker14}, \textit{PLATO} \citep{rauer14} and \textit{WFIRST} \citep{spergel13}. Each of these missions will provide high-precision, space-based photometry suitable for asteroseismology, with the expected number of detections of solar-like oscillations exceeding several million stars (see Fig.~\ref{fig:future}). Combined with ground-based efforts such as the SONG network \citep{grundahl08}, the key synergies between asteroseismology and exoplanet science are expected to be:

\begin{figure}[t]
\begin{center}
\resizebox{\hsize}{!}{\includegraphics{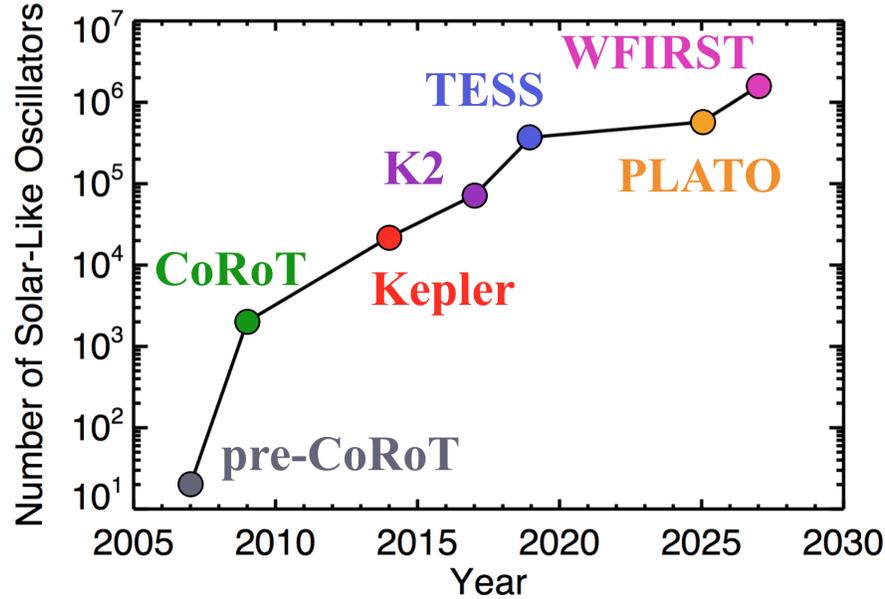}}
\caption{Number of stars with detected solar-like oscillations as a function of time. The approximate projected yield for current and future missions is $5\times10^{4}$ for K2 \citep[based on extrapolating classifications by][]{huber16}, $3\times10^{5}$ for \textit{TESS} (assuming detections in all red-clump stars down to $I \approx 10$\,mag with 27 days of data), $2\times10^{5}$ for \textit{PLATO} (assuming a similar red-giant fraction to \textit{Kepler}), and $10^{6}$ for \textit{WFIRST} \citep{gould15}. Note that $>$\,90\% of all detections are expected to be evolved stars, and \textit{PLATO} will by far contribute the most detections for dwarfs and subgiants \citep[$\approx$\,80,000 stars;][]{rauer14}.}
\label{fig:future}
\end{center}
\end{figure}

\begin{figure}[t]
\begin{center}
\resizebox{\hsize}{!}{\includegraphics{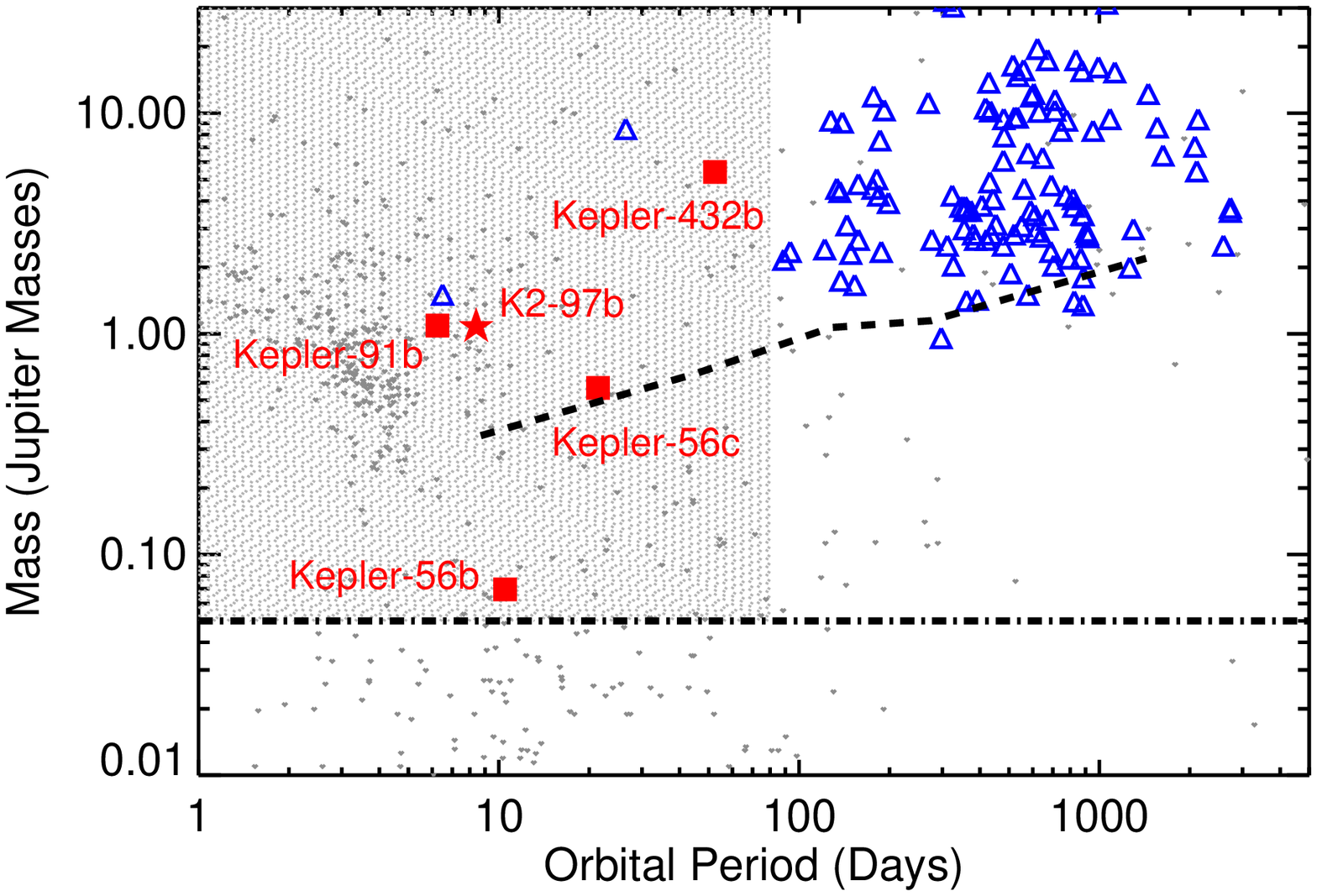}}
\caption{Mass versus orbital period for exoplanets orbiting 
evolved stars ($R_{\star}>3.5\,{\rm R}_{\odot}$, $T_{\rm eff} < 5500$\,K) detected with transits (red 
symbols) and radial velocities (blue triangles). Transiting systems are restricted to stars which are precisely characterized using asteroseismology.
The dashed line shows the median RV detection limit for mean masses given by \citet{bowler10}. 
The dashed-dotted line marks the mass of Neptune as an approximate K2 detection limit 
($R_{\rm P} \gtrsim 0.5 \, R_{\rm J}$), and the gray shaded area is probed by the K2 ``Giants Orbiting Giants Program''. Adapted from \citet{huber15b}.}
\label{fig:rgplanets}
\end{center}
\end{figure}

\begin{itemize}

\item \textbf{Densities and ages of sub-Neptune-size planets transiting asteroseismic solar-type stars:} In addition to \textit{TESS}, ground-based radial velocities obtained by SONG will play an important role in characterizing stars hosting small exoplanets using asteroseismology. Such systems will provide the best opportunity to precisely study the composition diversity of sub-Neptunes by constraining host star radii and masses to a few percent. Importantly, \textit{Gaia} parallaxes alone will not reach comparable precision due to model-dependent uncertainties such as bolometric corrections and reddening. The asteroseismology-exoplanet synergy is a core component of the \textit{PLATO} mission, which will extend the reach of asteroseismology to characterize radii, masses and ages of solar-type stars with small, transiting planets in the habitable zone.

\vspace{0.2cm}
\item \textbf{Gas-giant planets orbiting asteroseismic evolved stars:} Evolved stars provide an evolutionary ``sweet spot'' in which light curves with moderate cadence (such as the 30-minute sampling provided by \textit{Kepler}/K2 long-cadence data) can be used to detect transits and stellar oscillations simultaneously. Detections by \textit{Kepler} and the ``Giants Orbiting Giants Program'' with the K2 Mission \citep[see Fig.~\ref{fig:rgplanets};][]{huber15b} have demonstrated that these planets can be used to address key questions in exoplanetary science such as the effects of host star evolution on the radius inflation of hot Jupiters \citep{grunblatt16}. A particularly promising possibility to extend this synergy are full-frame images obtained by the \textit{TESS} mission, which are expected to yield several hundred asteroseismic exoplanet-host stars \citep{campante16}. Preliminary simulations have shown that low-luminosity RGB stars in the ecliptic poles with 1 year coverage can also be used to measure rotational splittings, and hence extend the study of exoplanet obliquities of systems similar to Kepler-56 (see Sect.~\ref{sec:obl}).

\vspace{0.2cm}
\item \textbf{Planets orbiting pulsating A stars:} Near-diffraction-limited, infrared adaptive-optics imaging instruments such as GPI \citep{macintosh08}, SPHERE \citep{beuzit08} and SCExAO \citep{guyon10} will soon provide an increasing number of directly imaged planets orbiting young stars, including pulsating A stars such as HR~8799 \citep{zerbi99,marois08}. A common limitation for interpreting these discoveries is their unknown age, which is needed to determine whether the detected substellar companions are indeed planets. While mode identification in $\delta$\,Scuti and $\gamma$\,Doradus stars is still challenging, the extension of the asteroseismology-exoplanet synergy to these systems will undoubtably become more important over the coming decades. Extended photometric monitoring provided over several years by \textit{PLATO} may also provide future opportunities to detect planets in wide orbits around A stars using pulsation frequency shifts induced by the planet \citep{murphy16}.
\end{itemize}

The above list is by no means complete, and further synergies beyond those discussed here will certainly be explored. With the wealth of data from ground-based and space-based facilities there is little doubt that the exciting and fruitful synergy between asteroseismology and exoplanetary science will continue to grow over the coming decades.

\begin{acknowledgement}
Many thanks to Tiago Campante, M\'ario Monteiro and all other organizers for a fantastic Summer School in Faial; and to Simon Albrecht, John Brewer, Vincent Van Eylen, Mikkel Lund, and Josh Winn for providing figures for this review. Financial support was provided by NASA grant NNX14AB92G and the Australian Research Council's Discovery Projects funding scheme (project number DEI40101364).
\end{acknowledgement}

\bibliographystyle{apj}
\bibliography{references}

\newcommand{\SortNoop}[1]{}
\begin{thebibliography}{}
\expandafter\ifx\csname natexlab\endcsname\relax\def\natexlab#1{#1}\fi

\bibitem[{{Adibekyan} {et~al.}(2014){Adibekyan}, {Gonz{\'a}lez Hern{\'a}ndez},
  {Delgado Mena}, {Sousa}, {Santos}, {Israelian}, {Figueira}, \& {Bertran de
  Lis}}]{adibekyan14}
{Adibekyan}, V.~Z., {Gonz{\'a}lez Hern{\'a}ndez}, J.~I., {Delgado Mena}, E.,
  {et~al.} 2014, \aap, 564, L15

\bibitem[{{Albrecht} {et~al.}(2013){Albrecht}, {Winn}, {Marcy}, {Howard},
  {Isaacson}, \& {Johnson}}]{albrecht13}
{Albrecht}, S., {Winn}, J.~N., {Marcy}, G.~W., {et~al.} 2013, \apj, 771, 11

\bibitem[{{Albrecht} {et~al.}(2012){Albrecht}, {Winn}, {Johnson}, {Howard},
  {Marcy}, {Butler}, {Arriagada}, {Crane}, {Shectman}, {Thompson}, {Hirano},
  {Bakos}, \& {Hartman}}]{albrecht12}
{Albrecht}, S., {Winn}, J.~N., {Johnson}, J.~A., {et~al.} 2012, \apj, 757, 18

\bibitem[{{Baglin} {et~al.}(2009){Baglin}, {Auvergne}, {Barge}, {Deleuil},
  {Michel}, \& {The CoRoT Exoplanet Science Team}}]{baglin09}
{Baglin}, A., {Auvergne}, M., {Barge}, P., {et~al.} 2009, in IAU Symposium,
  Vol. 253, IAU Symposium, 71--81

\bibitem[{{Ballard} {et~al.}(2014){Ballard}, {Chaplin}, {Charbonneau},
  {D{\'e}sert}, {Fressin}, {Zeng}, {Werner}, {Davies}, {Silva Aguirre}, {Basu},
  {Christensen-Dalsgaard}, {Metcalfe}, {Stello}, {Bedding}, {Campante},
  {Handberg}, {Karoff}, {Elsworth}, {Gilliland}, {Hekker}, {Huber}, {Kawaler},
  {Kjeldsen}, {Lund}, \& {Lundkvist}}]{ballard14}
{Ballard}, S., {Chaplin}, W.~J., {Charbonneau}, D., {et~al.} 2014, \apj, 790,
  12

\bibitem[{{Ballot} {et~al.}(2011){Ballot}, {Gizon}, {Samadi}, {Vauclair},
  {Benomar}, {Bruntt}, {Mosser}, {Stahn}, {Verner}, {Campante},
  {Garc{\'{\i}}a}, {Mathur}, {Salabert}, {Gaulme}, {R{\'e}gulo}, {Roxburgh},
  {Appourchaux}, {Baudin}, {Catala}, {Chaplin}, {Deheuvels}, {Michel}, {Bazot},
  {Creevey}, {Dolez}, {Elsworth}, {Sato}, {Vauclair}, {Auvergne}, \&
  {Baglin}}]{ballot11b}
{Ballot}, J., {Gizon}, L., {Samadi}, R., {et~al.} 2011, \aap, 530, A97

\bibitem[{{Barclay} {et~al.}(2012){Barclay}, {Huber}, {Rowe}, {Fortney},
  {Morley}, {Quintana}, {Fabrycky}, {Barentsen}, {Bloemen}, {Christiansen},
  {Demory}, {Fulton}, {Jenkins}, {Mullally}, {Ragozzine}, {Seader}, {Shporer},
  {Tenenbaum}, \& {Thompson}}]{barclay12}
{Barclay}, T., {Huber}, D., {Rowe}, J.~F., {et~al.} 2012, \apj, 761, 53

\bibitem[{{Barclay} {et~al.}(2013){Barclay}, {Burke}, {Howell}, {Rowe},
  {Huber}, {Isaacson}, {Jenkins}, {Kolbl}, {Marcy}, {Quintana}, {Still},
  {Twicken}, {Bryson}, {Borucki}, {Caldwell}, {Ciardi}, {Clarke},
  {Christiansen}, {Coughlin}, {Fischer}, {Li}, {Haas}, {Hunter}, {Lissauer},
  {Mullally}, {Sabale}, {Seader}, {Smith}, {Tenenbaum}, {Kamal Uddin}, \&
  {Thompson}}]{barclay13}
{Barclay}, T., {Burke}, C.~J., {Howell}, S.~B., {et~al.} 2013, \apj, 768, 101

\bibitem[{{Batygin} {et~al.}(2016){Batygin}, {Bodenheimer}, \&
  {Laughlin}}]{batygin16}
{Batygin}, K., {Bodenheimer}, P.~H., \& {Laughlin}, G.~P. 2016, \apj, 829, 114

\bibitem[{{Bazot} {et~al.}(2005){Bazot}, {Vauclair}, {Bouchy}, \&
  {Santos}}]{bazot05}
{Bazot}, M., {Vauclair}, S., {Bouchy}, F., \& {Santos}, N.~C. 2005, \aap, 440,
  615

\bibitem[{{Benomar} {et~al.}(2014){Benomar}, {Masuda}, {Shibahashi}, \&
  {Suto}}]{benomar14}
{Benomar}, O., {Masuda}, K., {Shibahashi}, H., \& {Suto}, Y. 2014, \pasj,
  arXiv:1407.7332

\bibitem[{{Beuzit} {et~al.}(2008){Beuzit}, {Feldt}, {Dohlen}, {Mouillet},
  {Puget}, {Wildi}, {Abe}, {Antichi}, {Baruffolo}, {Baudoz}, {Boccaletti},
  {Carbillet}, {Charton}, {Claudi}, {Downing}, {Fabron}, {Feautrier},
  {Fedrigo}, {Fusco}, {Gach}, {Gratton}, {Henning}, {Hubin}, {Joos}, {Kasper},
  {Langlois}, {Lenzen}, {Moutou}, {Pavlov}, {Petit}, {Pragt}, {Rabou}, {Rigal},
  {Roelfsema}, {Rousset}, {Saisse}, {Schmid}, {Stadler}, {Thalmann}, {Turatto},
  {Udry}, {Vakili}, \& {Waters}}]{beuzit08}
{Beuzit}, J.-L., {Feldt}, M., {Dohlen}, K., {et~al.} 2008, in \procspie, Vol.
  7014, Ground-based and Airborne Instrumentation for Astronomy II, 701418

\bibitem[{{Borucki} {et~al.}(2010){Borucki}, {Koch}, {Basri}, {Batalha},
  {Brown}, {Caldwell}, {Caldwell}, {Christensen-Dalsgaard}, {Cochran},
  {DeVore}, {Dunham}, {Dupree}, {Gautier}, {Geary}, {Gilliland}, {Gould},
  {Howell}, {Jenkins}, {Kondo}, {Latham}, {Marcy}, {Meibom}, {Kjeldsen},
  {Lissauer}, {Monet}, {Morrison}, {Sasselov}, {Tarter}, {Boss}, {Brownlee},
  {Owen}, {Buzasi}, {Charbonneau}, {Doyle}, {Fortney}, {Ford}, {Holman},
  {Seager}, {Steffen}, {Welsh}, {Rowe}, {Anderson}, {Buchhave}, {Ciardi},
  {Walkowicz}, {Sherry}, {Horch}, {Isaacson}, {Everett}, {Fischer}, {Torres},
  {Johnson}, {Endl}, {MacQueen}, {Bryson}, {Dotson}, {Haas}, {Kolodziejczak},
  {Van Cleve}, {Chandrasekaran}, {Twicken}, {Quintana}, {Clarke}, {Allen},
  {Li}, {Wu}, {Tenenbaum}, {Verner}, {Bruhweiler}, {Barnes}, \&
  {Prsa}}]{borucki10}
{Borucki}, W.~J., {Koch}, D., {Basri}, G., {et~al.} 2010, Science, 327, 977

\bibitem[{{Borucki} {et~al.}(2013){Borucki}, {Agol}, {Fressin}, {Kaltenegger},
  {Rowe}, {Isaacson}, {Fischer}, {Batalha}, {Lissauer}, {Marcy}, {Fabrycky},
  {D{\'e}sert}, {Bryson}, {Barclay}, {Bastien}, {Boss}, {Brugamyer},
  {Buchhave}, {Burke}, {Caldwell}, {Carter}, {Charbonneau}, {Crepp},
  {Christensen-Dalsgaard}, {Christiansen}, {Ciardi}, {Cochran}, {DeVore},
  {Doyle}, {Dupree}, {Endl}, {Everett}, {Ford}, {Fortney}, {Gautier}, {Geary},
  {Gould}, {Haas}, {Henze}, {Howard}, {Howell}, {Huber}, {Jenkins}, {Kjeldsen},
  {Kolbl}, {Kolodziejczak}, {Latham}, {Lee}, {Lopez}, {Mullally}, {Orosz},
  {Prsa}, {Quintana}, {Sanchis-Ojeda}, {Sasselov}, {Seader}, {Shporer},
  {Steffen}, {Still}, {Tenenbaum}, {Thompson}, {Torres}, {Twicken}, {Welsh}, \&
  {Winn}}]{borucki13}
{Borucki}, W.~J., {Agol}, E., {Fressin}, F., {et~al.} 2013, Science, 340, 587

\bibitem[{{Bouchy} {et~al.}(2005){Bouchy}, {Bazot}, {Santos}, {Vauclair}, \&
  {Sosnowska}}]{bouchy05}
{Bouchy}, F., {Bazot}, M., {Santos}, N.~C., {Vauclair}, S., \& {Sosnowska}, D.
  2005, \aap, 440, 609

\bibitem[{{Bowler} {et~al.}(2010){Bowler}, {Johnson}, {Marcy}, {Henry}, {Peek},
  {Fischer}, {Clubb}, {Liu}, {Reffert}, {Schwab}, \& {Lowe}}]{bowler10}
{Bowler}, B.~P., {Johnson}, J.~A., {Marcy}, G.~W., {et~al.} 2010, \apj, 709,
  396

\bibitem[{{Brewer} {et~al.}(2015){Brewer}, {Fischer}, {Basu}, {Valenti}, \&
  {Piskunov}}]{brewer15}
{Brewer}, J.~M., {Fischer}, D.~A., {Basu}, S., {Valenti}, J.~A., \& {Piskunov},
  N. 2015, \apj, 805, 126

\bibitem[{{Bruntt} {et~al.}(2010){Bruntt}, {Bedding}, {Quirion}, {Lo Curto},
  {Carrier}, {Smalley}, {Dall}, {Arentoft}, {Bazot}, \& {Butler}}]{bruntt10}
{Bruntt}, H., {Bedding}, T.~R., {Quirion}, P.-O., {et~al.} 2010, \mnras, 405,
  1907

\bibitem[{{Buchhave} {et~al.}(2012){Buchhave}, {Latham}, {Johansen},
  {Bizzarro}, {Torres}, {Rowe}, {Batalha}, {Borucki}, {Brugamyer}, {Caldwell},
  {Bryson}, {Ciardi}, {Cochran}, {Endl}, {Esquerdo}, {Ford}, {Geary},
  {Gilliland}, {Hansen}, {Isaacson}, {Laird}, {Lucas}, {Marcy}, {Morse},
  {Robertson}, {Shporer}, {Stefanik}, {Still}, \& {Quinn}}]{buchhave12}
{Buchhave}, L.~A., {Latham}, D.~W., {Johansen}, A., {et~al.} 2012, \nat, 486,
  375

\bibitem[{{Burke} {et~al.}(2015){Burke}, {Christiansen}, {Mullally}, {Seader},
  {Huber}, {Rowe}, {Coughlin}, {Thompson}, {Catanzarite}, {Clarke}, {Morton},
  {Caldwell}, {Bryson}, {Haas}, {Batalha}, {Jenkins}, {Tenenbaum}, {Twicken},
  {Li}, {Quintana}, {Barclay}, {Henze}, {Borucki}, {Howell}, \&
  {Still}}]{burke15}
{Burke}, C.~J., {Christiansen}, J.~L., {Mullally}, F., {et~al.} 2015, \apj,
  809, 8

\bibitem[{{Campante} {et~al.}(2015){Campante}, {Barclay}, {Swift}, {Huber},
  {Adibekyan}, {Cochran}, {Burke}, {Isaacson}, {Quintana}, {Davies}, {Silva
  Aguirre}, {Ragozzine}, {Riddle}, {Baranec}, {Basu}, {Chaplin},
  {Christensen-Dalsgaard}, {Metcalfe}, {Bedding}, {Handberg}, {Stello},
  {Brewer}, {Hekker}, {Karoff}, {Kolbl}, {Law}, {Lundkvist}, {Miglio}, {Rowe},
  {Santos}, {Van Laerhoven}, {Arentoft}, {Elsworth}, {Fischer}, {Kawaler},
  {Kjeldsen}, {Lund}, {Marcy}, {Sousa}, {Sozzetti}, \& {White}}]{campante15}
{Campante}, T.~L., {Barclay}, T., {Swift}, J.~J., {et~al.} 2015, \apj, 799, 170

\bibitem[{{Campante} {et~al.}(2016){Campante}, {Schofield}, {Kuszlewicz},
  {Bouma}, {Chaplin}, {Huber}, {Christensen-Dalsgaard}, {Kjeldsen}, {Bossini},
  {North}, {Appourchaux}, {Latham}, {Pepper}, {Ricker}, {Stassun},
  {Vanderspek}, \& {Winn}}]{campante16}
{Campante}, T.~L., {Schofield}, M., {Kuszlewicz}, J.~S., {et~al.} 2016, \apj,
  830, 138

\bibitem[{{Carter} {et~al.}(2012){Carter}, {Agol}, {Chaplin}, {Basu},
  {Bedding}, {Buchhave}, {Christensen-Dalsgaard}, {Deck}, {Elsworth},
  {Fabrycky}, {Ford}, {Fortney}, {Hale}, {Handberg}, {Hekker}, {Holman},
  {Huber}, {Karoff}, {Kawaler}, {Kjeldsen}, {Lissauer}, {Lopez}, {Lund},
  {Lundkvist}, {Metcalfe}, {Miglio}, {Rogers}, {Stello}, {Borucki}, {Bryson},
  {Christiansen}, {Cochran}, {Geary}, {Gilliland}, {Haas}, {Hall}, {Howard},
  {Jenkins}, {Klaus}, {Koch}, {Latham}, {MacQueen}, {Sasselov}, {Steffen},
  {Twicken}, \& {Winn}}]{carter12}
{Carter}, J.~A., {Agol}, E., {Chaplin}, W.~J., {et~al.} 2012, Science, 337, 556

\bibitem[{{Chaplin} {et~al.}(2013){Chaplin}, {Sanchis-Ojeda}, {Campante},
  {Handberg}, {Stello}, {Winn}, {Basu}, {Christensen-Dalsgaard}, {Davies},
  {Metcalfe}, {Buchhave}, {Fischer}, {Bedding}, {Cochran}, {Elsworth},
  {Gilliland}, {Hekker}, {Huber}, {Isaacson}, {Karoff}, {Kawaler}, {Kjeldsen},
  {Latham}, {Lund}, {Lundkvist}, {Marcy}, {Miglio}, {Barclay}, \&
  {Lissauer}}]{chaplin13c}
{Chaplin}, W.~J., {Sanchis-Ojeda}, R., {Campante}, T.~L., {et~al.} 2013, \apj,
  766, 101

\bibitem[{{Chaplin} {et~al.}(2014){Chaplin}, {Basu}, {Huber}, {Serenelli},
  {Casagrande}, {Silva Aguirre}, {Ball}, {Creevey}, {Gizon}, {Handberg},
  {Karoff}, {Lutz}, {Marques}, {Miglio}, {Stello}, {Suran}, {Pricopi},
  {Metcalfe}, {Monteiro}, {Molenda-{\.Z}akowicz}, {Appourchaux},
  {Christensen-Dalsgaard}, {Elsworth}, {Garc{\'{\i}}a}, {Houdek}, {Kjeldsen},
  {Bonanno}, {Campante}, {Corsaro}, {Gaulme}, {Hekker}, {Mathur}, {Mosser},
  {R{\'e}gulo}, \& {Salabert}}]{chaplin13}
{Chaplin}, W.~J., {Basu}, S., {Huber}, D., {et~al.} 2014, \apjs, 210, 1

\bibitem[{{Chatterjee} {et~al.}(2008){Chatterjee}, {Ford}, {Matsumura}, \&
  {Rasio}}]{chatterjee08}
{Chatterjee}, S., {Ford}, E.~B., {Matsumura}, S., \& {Rasio}, F.~A. 2008, \apj,
  686, 580

\bibitem[{{Christensen-Dalsgaard} {et~al.}(2010){Christensen-Dalsgaard},
  {Kjeldsen}, {Brown}, {Gilliland}, {Arentoft}, {Frandsen}, {Quirion},
  {Borucki}, {Koch}, \& {Jenkins}}]{cd10}
{Christensen-Dalsgaard}, J., {Kjeldsen}, H., {Brown}, T.~M., {et~al.} 2010,
  \apjl, 713, L164

\bibitem[{{Davies} {et~al.}(2016){Davies}, {Silva Aguirre}, {Bedding},
  {Handberg}, {Lund}, {Chaplin}, {Huber}, {White}, {Benomar}, {Hekker}, {Basu},
  {Campante}, {Christensen-Dalsgaard}, {Elsworth}, {Karoff}, {Kjeldsen},
  {Lundkvist}, {Metcalfe}, \& {Stello}}]{davies15}
{Davies}, G.~R., {Silva Aguirre}, V., {Bedding}, T.~R., {et~al.} 2016, \mnras,
  456, 2183

\bibitem[{{Dong} \& {Zhu}(2013)}]{dong13b}
{Dong}, S., \& {Zhu}, Z. 2013, \apj, 778, 53

\bibitem[{{Dupuy} {et~al.}(2016){Dupuy}, {Kratter}, {Kraus}, {Isaacson},
  {Mann}, {Ireland}, {Howard}, \& {Huber}}]{dupuy16}
{Dupuy}, T.~J., {Kratter}, K.~M., {Kraus}, A.~L., {et~al.} 2016, \apj, 817, 80

\bibitem[{{Fabrycky} \& {Tremaine}(2007)}]{fabrycky07}
{Fabrycky}, D., \& {Tremaine}, S. 2007, \apj, 669, 1298

\bibitem[{{Fabrycky} \& {Winn}(2009)}]{fabrycky09}
{Fabrycky}, D.~C., \& {Winn}, J.~N. 2009, \apj, 696, 1230

\bibitem[{{Fischer} \& {Valenti}(2005)}]{fischer05}
{Fischer}, D.~A., \& {Valenti}, J. 2005, \apj, 622, 1102

\bibitem[{{Frandsen} {et~al.}(2013){Frandsen}, {Lehmann}, {Hekker},
  {Southworth}, {Debosscher}, {Beck}, {Hartmann}, {Pigulski}, {Kopacki},
  {Ko{\l}aczkowski}, {St{\c e}{\'s}licki}, {Thygesen}, {Brogaard}, \&
  {Elsworth}}]{frandsen13}
{Frandsen}, S., {Lehmann}, H., {Hekker}, S., {et~al.} 2013, \aap, 556, A138

\bibitem[{{Gaulme} {et~al.}(2016){Gaulme}, {McKeever}, {Jackiewicz}, {Rawls},
  {Corsaro}, {Mosser}, {Southworth}, {Mahadevan}, {Bender}, \&
  {Deshpande}}]{gaulme16}
{Gaulme}, P., {McKeever}, J., {Jackiewicz}, J., {et~al.} 2016, \apj, 832, 121

\bibitem[{{Gilliland} {et~al.}(2011){Gilliland}, {McCullough}, {Nelan},
  {Brown}, {Charbonneau}, {Nutzman}, {Christensen-Dalsgaard}, \&
  {Kjeldsen}}]{gilliland11}
{Gilliland}, R.~L., {McCullough}, P.~R., {Nelan}, E.~P., {et~al.} 2011, \apj,
  726, 2

\bibitem[{{Gizon} \& {Solanki}(2003)}]{gizon03}
{Gizon}, L., \& {Solanki}, S.~K. 2003, \apj, 589, 1009

\bibitem[{{Gonzalez}(1997)}]{gonzalez97}
{Gonzalez}, G. 1997, \mnras, 285, 403

\bibitem[{{Gould} {et~al.}(2015){Gould}, {Huber}, {Penny}, \&
  {Stello}}]{gould15}
{Gould}, A., {Huber}, D., {Penny}, M., \& {Stello}, D. 2015, Journal of Korean
  Astronomical Society, 48, 93

\bibitem[{{Gratia} \& {Fabrycky}(2017)}]{gratia17}
{Gratia}, P., \& {Fabrycky}, D. 2017, \mnras, 464, 1709

\bibitem[{{Grunblatt} {et~al.}(2016){Grunblatt}, {Huber}, {Gaidos}, {Lopez},
  {Fulton}, {Vanderburg}, {Barclay}, {Fortney}, {Howard}, {Isaacson}, {Mann},
  {Petigura}, {Silva Aguirre}, \& {Sinukoff}}]{grunblatt16}
{Grunblatt}, S.~K., {Huber}, D., {Gaidos}, E.~J., {et~al.} 2016, \aj, 152, 185

\bibitem[{{Grundahl} {et~al.}(2008){Grundahl}, {Christensen-Dalsgaard},
  {Kjeldsen}, {Frandsen}, {Arentoft}, {Kjaergaard}, \&
  {J{\o}rgensen}}]{grundahl08}
{Grundahl}, F., {Christensen-Dalsgaard}, J., {Kjeldsen}, H., {et~al.} 2008, in
  IAU Symposium, Vol. 252, IAU Symposium, ed. {L.~Deng \& K.~L.~Chan}, 465--466

\bibitem[{{Guyon} {et~al.}(2010){Guyon}, {Martinache}, {Garrel}, {Vogt},
  {Yokochi}, \& {Yoshikawa}}]{guyon10}
{Guyon}, O., {Martinache}, F., {Garrel}, V., {et~al.} 2010, in \procspie, Vol.
  7736, Adaptive Optics Systems II, 773624

\bibitem[{{Hadden} \& {Lithwick}(2014)}]{hadden14}
{Hadden}, S., \& {Lithwick}, Y. 2014, \apj, 787, 80

\bibitem[{{Howard} {et~al.}(2012){Howard}, {Marcy}, {Bryson}, {Jenkins},
  {Rowe}, {Batalha}, {Borucki}, {Koch}, {Dunham}, {Gautier}, {Van Cleve},
  {Cochran}, {Latham}, {Lissauer}, {Torres}, {Brown}, {Gilliland}, {Buchhave},
  {Caldwell}, {Christensen-Dalsgaard}, {Ciardi}, {Fressin}, {Haas}, {Howell},
  {Kjeldsen}, {Seager}, {Rogers}, {Sasselov}, {Steffen}, {Basri},
  {Charbonneau}, {Christiansen}, {Clarke}, {Dupree}, {Fabrycky}, {Fischer},
  {Ford}, {Fortney}, {Tarter}, {Girouard}, {Holman}, {Johnson}, {Klaus},
  {Machalek}, {Moorhead}, {Morehead}, {Ragozzine}, {Tenenbaum}, {Twicken},
  {Quinn}, {Isaacson}, {Shporer}, {Lucas}, {Walkowicz}, {Welsh}, {Boss},
  {Devore}, {Gould}, {Smith}, {Morris}, {Prsa}, {Morton}, {Still}, {Thompson},
  {Mullally}, {Endl}, \& {MacQueen}}]{howard11}
{Howard}, A.~W., {Marcy}, G.~W., {Bryson}, S.~T., {et~al.} 2012, \apjs, 201, 15

\bibitem[{{Huber}(2015{\natexlab{a}})}]{huber14b}
{Huber}, D. 2015{\natexlab{a}}, in Astrophysics and Space Science Library, Vol.
  408, Giants of Eclipse: The zeta Aurigae Stars and Other Binary Systems, 169

\bibitem[{{Huber}(2015{\natexlab{b}})}]{huber15b}
{Huber}, D. 2015{\natexlab{b}}, ArXiv e-prints, arXiv:1511.07441

\bibitem[{{Huber} {et~al.}(2013{\natexlab{a}}){Huber}, {Chaplin},
  {Christensen-Dalsgaard}, {Gilliland}, {Kjeldsen}, {Buchhave}, {Fischer},
  {Lissauer}, {Rowe}, {Sanchis-Ojeda}, {Basu}, {Handberg}, {Hekker}, {Howard},
  {Isaacson}, {Karoff}, {Latham}, {Lund}, {Lundkvist}, {Marcy}, {Miglio},
  {Silva Aguirre}, {Stello}, {Arentoft}, {Barclay}, {Bedding}, {Burke},
  {Christiansen}, {Elsworth}, {Haas}, {Kawaler}, {Metcalfe}, {Mullally}, \&
  {Thompson}}]{huber13}
{Huber}, D., {Chaplin}, W.~J., {Christensen-Dalsgaard}, J., {et~al.}
  2013{\natexlab{a}}, \apj, 767, 127

\bibitem[{{Huber} {et~al.}(2013{\natexlab{b}}){Huber}, {Carter}, {Barbieri},
  {Miglio}, {Deck}, {Fabrycky}, {Montet}, {Buchhave}, {Chaplin}, {Hekker},
  {Montalb{\'a}n}, {Sanchis-Ojeda}, {Basu}, {Bedding}, {Campante},
  {Christensen-Dalsgaard}, {Elsworth}, {Stello}, {Arentoft}, {Ford},
  {Gilliland}, {Handberg}, {Howard}, {Isaacson}, {Johnson}, {Karoff},
  {Kawaler}, {Kjeldsen}, {Latham}, {Lund}, {Lundkvist}, {Marcy}, {Metcalfe},
  {Silva Aguirre}, \& {Winn}}]{huber13b}
{Huber}, D., {Carter}, J.~A., {Barbieri}, M., {et~al.} 2013{\natexlab{b}},
  Science, 342, 331

\bibitem[{{Huber} {et~al.}(2016){Huber}, {Bryson}, {Haas}, {Barclay},
  {Barentsen}, {Howell}, {Sharma}, {Stello}, \& {Thompson}}]{huber16}
{Huber}, D., {Bryson}, S.~T., {Haas}, M.~R., {et~al.} 2016, \apjs, 224, 2

\bibitem[{{Jenkins} {et~al.}(2015){Jenkins}, {Twicken}, {Batalha}, {Caldwell},
  {Cochran}, {Endl}, {Latham}, {Esquerdo}, {Seader}, {Bieryla}, {Petigura},
  {Ciardi}, {Marcy}, {Isaacson}, {Huber}, {Rowe}, {Torres}, {Bryson},
  {Buchhave}, {Ramirez}, {Wolfgang}, {Li}, {Campbell}, {Tenenbaum},
  {Sanderfer}, {Henze}, {Catanzarite}, {Gilliland}, \& {Borucki}}]{jenkins15}
{Jenkins}, J.~M., {Twicken}, J.~D., {Batalha}, N.~M., {et~al.} 2015, \aj, 150,
  56

\bibitem[{{Johnson} {et~al.}(2009){Johnson}, {Winn}, {Albrecht}, {Howard},
  {Marcy}, \& {Gazak}}]{johnson09}
{Johnson}, J.~A., {Winn}, J.~N., {Albrecht}, S., {et~al.} 2009, \pasp, 121,
  1104

\bibitem[{{Kallinger} {et~al.}(2010){Kallinger}, {Mosser}, {Hekker}, {Huber},
  {Stello}, {Mathur}, {Basu}, {Bedding}, {Chaplin}, {De Ridder}, {Elsworth},
  {Frandsen}, {Garc{\'{\i}}a}, {Gruberbauer}, {Matthews}, {Borucki}, {Bruntt},
  {Christensen-Dalsgaard}, {Gilliland}, {Kjeldsen}, \& {Koch}}]{kallinger10}
{Kallinger}, T., {Mosser}, B., {Hekker}, S., {et~al.} 2010, \aap, 522, A1

\bibitem[{{Kane}(2014)}]{kane14}
{Kane}, S.~R. 2014, \apj, 782, 111

\bibitem[{{Kipping}(2010)}]{kipping10}
{Kipping}, D.~M. 2010, \mnras, 407, 301

\bibitem[{{Lebreton} \& {Goupil}(2014)}]{lebreton14}
{Lebreton}, Y., \& {Goupil}, M.~J. 2014, \aap, 569, A21

\bibitem[{{Li} {et~al.}(2014){Li}, {Naoz}, {Valsecchi}, {Johnson}, \&
  {Rasio}}]{li14}
{Li}, G., {Naoz}, S., {Valsecchi}, F., {Johnson}, J.~A., \& {Rasio}, F.~A.
  2014, \apj, 794, 131

\bibitem[{{Lin} {et~al.}(1996){Lin}, {Bodenheimer}, \& {Richardson}}]{lin96}
{Lin}, D.~N.~C., {Bodenheimer}, P., \& {Richardson}, D.~C. 1996, \nat, 380, 606

\bibitem[{{Lissauer} {et~al.}(2012){Lissauer}, {Marcy}, {Rowe}, {Bryson},
  {Adams}, {Buchhave}, {Ciardi}, {Cochran}, {Fabrycky}, {Ford}, {Fressin},
  {Geary}, {Gilliland}, {Holman}, {Howell}, {Jenkins}, {Kinemuchi}, {Koch},
  {Morehead}, {Ragozzine}, {Seader}, {Tanenbaum}, {Torres}, \&
  {Twicken}}]{lissauer12}
{Lissauer}, J.~J., {Marcy}, G.~W., {Rowe}, J.~F., {et~al.} 2012, \apj, 750, 112

\bibitem[{{Lithwick} {et~al.}(2012){Lithwick}, {Xie}, \& {Wu}}]{lithwick12}
{Lithwick}, Y., {Xie}, J., \& {Wu}, Y. 2012, \apj, 761, 122

\bibitem[{{Lund} {et~al.}(2014){Lund}, {Lundkvist}, {Silva Aguirre}, {Houdek},
  {Casagrande}, {Van Eylen}, {Campante}, {Karoff}, {Kjeldsen}, {Albrecht},
  {Chaplin}, {Nielsen}, {Degroote}, {Davies}, \& {Handberg}}]{lund14}
{Lund}, M.~N., {Lundkvist}, M., {Silva Aguirre}, V., {et~al.} 2014, \aap, 570,
  A54

\bibitem[{{Lundkvist} {et~al.}(2016){Lundkvist}, {Kjeldsen}, {Albrecht},
  {Davies}, {Basu}, {Huber}, {Justesen}, {Karoff}, {Silva Aguirre}, {Van
  Eylen}, {Vang}, {Arentoft}, {Barclay}, {Bedding}, {Campante}, {Chaplin},
  {Christensen-Dalsgaard}, {Elsworth}, {Gilliland}, {Handberg}, {Hekker},
  {Kawaler}, {Lund}, {Metcalfe}, {Miglio}, {Rowe}, {Stello}, {Tingley}, \&
  {White}}]{lundkvist16}
{Lundkvist}, M.~S., {Kjeldsen}, H., {Albrecht}, S., {et~al.} 2016, Nature
  Communications, 7, 11201

\bibitem[{{Macintosh} {et~al.}(2008){Macintosh}, {Graham}, {Palmer}, {Doyon},
  {Dunn}, {Gavel}, {Larkin}, {Oppenheimer}, {Saddlemyer}, {Sivaramakrishnan},
  {Wallace}, {Bauman}, {Erickson}, {Marois}, {Poyneer}, \&
  {Soummer}}]{macintosh08}
{Macintosh}, B.~A., {Graham}, J.~R., {Palmer}, D.~W., {et~al.} 2008, in
  \procspie, Vol. 7015, Adaptive Optics Systems, 701518

\bibitem[{{Marois} {et~al.}(2008){Marois}, {Macintosh}, {Barman}, {Zuckerman},
  {Song}, {Patience}, {Lafreni{\`e}re}, \& {Doyon}}]{marois08}
{Marois}, C., {Macintosh}, B., {Barman}, T., {et~al.} 2008, Science, 322, 1348

\bibitem[{{Matsakos} \& {K{\"o}nigl}(2017)}]{matsakos17}
{Matsakos}, T., \& {K{\"o}nigl}, A. 2017, \aj, 153, 60

\bibitem[{{Mel{\'e}ndez} {et~al.}(2009){Mel{\'e}ndez}, {Asplund}, {Gustafsson},
  \& {Yong}}]{melendez09}
{Mel{\'e}ndez}, J., {Asplund}, M., {Gustafsson}, B., \& {Yong}, D. 2009, \apjl,
  704, L66

\bibitem[{{Mortier} {et~al.}(2014){Mortier}, {Sousa}, {Adibekyan},
  {Brand{\~a}o}, \& {Santos}}]{mortier14}
{Mortier}, A., {Sousa}, S.~G., {Adibekyan}, V.~Z., {Brand{\~a}o}, I.~M., \&
  {Santos}, N.~C. 2014, \aap, 572, A95

\bibitem[{{Murphy} {et~al.}(2016){Murphy}, {Bedding}, \&
  {Shibahashi}}]{murphy16}
{Murphy}, S.~J., {Bedding}, T.~R., \& {Shibahashi}, H. 2016, \apjl, 827, L17

\bibitem[{{Nutzman} {et~al.}(2011){Nutzman}, {Gilliland}, {McCullough},
  {Charbonneau}, {Christensen-Dalsgaard}, {Kjeldsen}, {Nelan}, {Brown}, \&
  {Holman}}]{nutzman11}
{Nutzman}, P., {Gilliland}, R.~L., {McCullough}, P.~R., {et~al.} 2011, \apj,
  726, 3

\bibitem[{{Otor} {et~al.}(2016){Otor}, {Montet}, {Johnson}, {Charbonneau},
  {Collier-Cameron}, {Howard}, {Isaacson}, {Latham}, {Lopez-Morales}, {Lovis},
  {Mayor}, {Micela}, {Molinari}, {Pepe}, {Piotto}, {Phillips}, {Queloz},
  {Rice}, {Sasselov}, {S{\'e}gransan}, {Sozzetti}, {Udry}, \&
  {Watson}}]{otor16}
{Otor}, O.~J., {Montet}, B.~T., {Johnson}, J.~A., {et~al.} 2016, \aj, 152, 165

\bibitem[{{Petigura} {et~al.}(2013){Petigura}, {{\SortNoop{z}}Howard}, \&
  {Marcy}}]{petigura13b}
{Petigura}, E.~A., {{\SortNoop{z}}Howard}, A.~W., \& {Marcy}, G.~W. 2013, PNAS,
  110, 19175

\bibitem[{{Pietrinferni} {et~al.}(2004){Pietrinferni}, {Cassisi}, {Salaris}, \&
  {Castelli}}]{basti}
{Pietrinferni}, A., {Cassisi}, S., {Salaris}, M., \& {Castelli}, F. 2004, \apj,
  612, 168

\bibitem[{{Quinn} {et~al.}(2015){Quinn}, {White}, {Latham}, {Chaplin},
  {Handberg}, {Huber}, {Kipping}, {Payne}, {Jiang}, {Silva Aguirre}, {Stello},
  {Sliski}, {Ciardi}, {Buchhave}, {Bedding}, {Davies}, {Hekker}, {Kjeldsen},
  {Kuszlewicz}, {Everett}, {Howell}, {Basu}, {Campante},
  {Christensen-Dalsgaard}, {Elsworth}, {Karoff}, {Kawaler}, {Lund},
  {Lundkvist}, {Esquerdo}, {Calkins}, \& {Berlind}}]{quinn15}
{Quinn}, S.~N., {White}, T.~R., {Latham}, D.~W., {et~al.} 2015, \apj, 803, 49

\bibitem[{{Quintana} {et~al.}(2014){Quintana}, {Barclay}, {Raymond}, {Rowe},
  {Bolmont}, {Caldwell}, {Howell}, {Kane}, {Huber}, {Crepp}, {Lissauer},
  {Ciardi}, {Coughlin}, {Everett}, {Henze}, {Horch}, {Isaacson}, {Ford},
  {Adams}, {Still}, {Hunter}, {Quarles}, \& {Selsis}}]{quintana14}
{Quintana}, E.~V., {Barclay}, T., {Raymond}, S.~N., {et~al.} 2014, Science,
  344, 277

\bibitem[{{Ram{\'{\i}}rez} {et~al.}(2009){Ram{\'{\i}}rez}, {Mel{\'e}ndez}, \&
  {Asplund}}]{ramirez09}
{Ram{\'{\i}}rez}, I., {Mel{\'e}ndez}, J., \& {Asplund}, M. 2009, \aap, 508, L17

\bibitem[{{Rauer} {et~al.}(2014){Rauer}, {Catala}, {Aerts}, {Appourchaux},
  {Benz}, {Brandeker}, {Christensen-Dalsgaard}, {Deleuil}, {Gizon}, {Goupil},
  {G{\"u}del}, {Janot-Pacheco}, {Mas-Hesse}, {Pagano}, {Piotto}, {Pollacco},
  {Santos}, {Smith}, {Su{\'a}rez}, {Szab{\'o}}, {Udry}, {Adibekyan}, {Alibert},
  {Almenara}, {Amaro-Seoane}, {Ammer-von Eiff}, {Asplund}, {Antonello},
  {Barnes}, {Baudin}, {Belkacem}, {Bergemann}, {Bihain}, {Birch}, {Bonfils},
  {Boisse}, {Bonomo}, {Borsa}, {Brand{\~a}o}, {Brocato}, {Brun}, {Burleigh},
  {Burston}, {Cabrera}, {Cassisi}, {Chaplin}, {Charpinet}, {Chiappini},
  {Church}, {Csizmadia}, {Cunha}, {Damasso}, {Davies}, {Deeg}, {D{\'{\i}}az},
  {Dreizler}, {Dreyer}, {Eggenberger}, {Ehrenreich}, {Eigm{\"u}ller},
  {Erikson}, {Farmer}, {Feltzing}, {Oliveira Fialho}, {Figueira}, {Forveille},
  {Fridlund}, {Garc{\'{\i}}a}, {Giommi}, {Giuffrida}, {Godolt}, {Gomes da
  Silva}, {Granzer}, {Grenfell}, {Grotsch-Noels}, {G{\"u}nther}, {Haswell},
  {Hatzes}, {H{\'e}brard}, {Hekker}, {Helled}, {Heng}, {Jenkins}, {Johansen},
  {Khodachenko}, {Kislyakova}, {Kley}, {Kolb}, {Krivova}, {Kupka}, {Lammer},
  {Lanza}, {Lebreton}, {Magrin}, {Marcos-Arenal}, {Marrese}, {Marques},
  {Martins}, {Mathis}, {Mathur}, {Messina}, {Miglio}, {Montalban}, {Montalto},
  {Monteiro}, {Moradi}, {Moravveji}, {Mordasini}, {Morel}, {Mortier},
  {Nascimbeni}, {Nelson}, {Nielsen}, {Noack}, {Norton}, {Ofir}, {Oshagh},
  {Ouazzani}, {P{\'a}pics}, {Parro}, {Petit}, {Plez}, {Poretti}, {Quirrenbach},
  {Ragazzoni}, {Raimondo}, {Rainer}, {Reese}, {Redmer}, {Reffert},
  {Rojas-Ayala}, {Roxburgh}, {Salmon}, {Santerne}, {Schneider}, {Schou},
  {Schuh}, {Schunker}, {Silva-Valio}, {Silvotti}, {Skillen}, {Snellen}, {Sohl},
  {Sousa}, {Sozzetti}, {Stello}, {Strassmeier}, {{\v S}vanda}, {Szab{\'o}},
  {Tkachenko}, {Valencia}, {Van Grootel}, {Vauclair}, {Ventura}, {Wagner},
  {Walton}, {Weingrill}, {Werner}, {Wheatley}, \& {Zwintz}}]{rauer14}
{Rauer}, H., {Catala}, C., {Aerts}, C., {et~al.} 2014, Experimental Astronomy,
  arXiv:1310.0696

\bibitem[{{Ricker} {et~al.}(2014){Ricker}, {Winn}, {Vanderspek}, {Latham},
  {Bakos}, {Bean}, {Berta-Thompson}, {Brown}, {Buchhave}, {Butler}, {Butler},
  {Chaplin}, {Charbonneau}, {Christensen-Dalsgaard}, {Clampin}, {Deming},
  {Doty}, {De Lee}, {Dressing}, {Dunham}, {Endl}, {Fressin}, {Ge}, {Henning},
  {Holman}, {Howard}, {Ida}, {Jenkins}, {Jernigan}, {Johnson}, {Kaltenegger},
  {Kawai}, {Kjeldsen}, {Laughlin}, {Levine}, {Lin}, {Lissauer}, {MacQueen},
  {Marcy}, {McCullough}, {Morton}, {Narita}, {Paegert}, {Palle}, {Pepe},
  {Pepper}, {Quirrenbach}, {Rinehart}, {Sasselov}, {Sato}, {Seager},
  {Sozzetti}, {Stassun}, {Sullivan}, {Szentgyorgyi}, {Torres}, {Udry}, \&
  {Villasenor}}]{ricker14}
{Ricker}, G.~R., {Winn}, J.~N., {Vanderspek}, R., {et~al.} 2014, in \procspie,
  Vol. 9143, Space Telescopes and Instrumentation 2014: Optical, Infrared, and
  Millimeter Wave, 914320

\bibitem[{{Rogers}(2015)}]{rogers15}
{Rogers}, L.~A. 2015, \apj, 801, 41

\bibitem[{{Sanchis-Ojeda} {et~al.}(2013){Sanchis-Ojeda}, {Winn}, {Marcy},
  {Howard}, {Isaacson}, {Johnson}, {Torres}, {Albrecht}, {Campante}, {Chaplin},
  {Davies}, {Lund}, {Carter}, {Dawson}, {Buchhave}, {Everett}, {Fischer},
  {Geary}, {Gilliland}, {Horch}, {Howell}, \& {Latham}}]{sanchis13}
{Sanchis-Ojeda}, R., {Winn}, J.~N., {Marcy}, G.~W., {et~al.} 2013, \apj, 775,
  54

\bibitem[{{Seager} \& {Mall{\'e}n-Ornelas}(2003)}]{seager03}
{Seager}, S., \& {Mall{\'e}n-Ornelas}, G. 2003, \apj, 585, 1038

\bibitem[{{Silva Aguirre} {et~al.}(2015){Silva Aguirre}, {Davies}, {Basu},
  {Christensen-Dalsgaard}, {Creevey}, {Metcalfe}, {Bedding}, {Casagrande},
  {Handberg}, {Lund}, {Nissen}, {Chaplin}, {Huber}, {Serenelli}, {Stello}, {Van
  Eylen}, {Campante}, {Elsworth}, {Gilliland}, {Hekker}, {Karoff}, {Kawaler},
  {Kjeldsen}, \& {Lundkvist}}]{silva15}
{Silva Aguirre}, V., {Davies}, G.~R., {Basu}, S., {et~al.} 2015, \mnras, 452,
  2127

\bibitem[{{Sliski} \& {Kipping}(2014)}]{sliski14}
{Sliski}, D.~H., \& {Kipping}, D.~M. 2014, \apj, 788, 148

\bibitem[{{Southworth}(2011)}]{southworth11}
{Southworth}, J. 2011, \mnras, 417, 2166

\bibitem[{{Southworth}(2012)}]{southworth12}
---. 2012, \mnras, 426, 1291

\bibitem[{{Spergel} {et~al.}(2013){Spergel}, {Gehrels}, {Breckinridge},
  {Donahue}, {Dressler}, {Gaudi}, {Greene}, {Guyon}, {Hirata}, {Kalirai},
  {Kasdin}, {Moos}, {Perlmutter}, {Postman}, {Rauscher}, {Rhodes}, {Wang},
  {Weinberg}, {Centrella}, {Traub}, {Baltay}, {Colbert}, {Bennett},
  {Kiessling}, {Macintosh}, {Merten}, {Mortonson}, {Penny}, {Rozo},
  {Savransky}, {Stapelfeldt}, {Zu}, {Baker}, {Cheng}, {Content}, {Dooley},
  {Foote}, {Goullioud}, {Grady}, {Jackson}, {Kruk}, {Levine}, {Melton},
  {Peddie}, {Ruffa}, \& {Shaklan}}]{spergel13}
{Spergel}, D., {Gehrels}, N., {Breckinridge}, J., {et~al.} 2013, ArXiv
  e-prints, arXiv:1305.5425

\bibitem[{{Steffen} {et~al.}(2012){Steffen}, {Fabrycky}, {Ford}, {Carter},
  {D{\'e}sert}, {Fressin}, {Holman}, {Lissauer}, {Moorhead}, {Rowe},
  {Ragozzine}, {Welsh}, {Batalha}, {Borucki}, {Buchhave}, {Bryson}, {Caldwell},
  {Charbonneau}, {Ciardi}, {Cochran}, {Endl}, {Everett}, {Gautier},
  {Gilliland}, {Girouard}, {Jenkins}, {Horch}, {Howell}, {Isaacson}, {Klaus},
  {Koch}, {Latham}, {Li}, {Lucas}, {MacQueen}, {Marcy}, {McCauliff}, {Middour},
  {Morris}, {Mullally}, {Quinn}, {Quintana}, {Shporer}, {Still}, {Tenenbaum},
  {Thompson}, {Twicken}, \& {Van Cleve}}]{steffen12}
{Steffen}, J.~H., {Fabrycky}, D.~C., {Ford}, E.~B., {et~al.} 2012, \mnras, 421,
  2342

\bibitem[{{Torres} {et~al.}(2012){Torres}, {Fischer}, {Sozzetti}, {Buchhave},
  {Winn}, {Holman}, \& {Carter}}]{torres12}
{Torres}, G., {Fischer}, D.~A., {Sozzetti}, A., {et~al.} 2012, \apj, 757, 161

\bibitem[{{Van Eylen} \& {Albrecht}(2015)}]{vaneylen15}
{Van Eylen}, V., \& {Albrecht}, S. 2015, \apj, 808, 126

\bibitem[{{Van Eylen} {et~al.}(2014){Van Eylen}, {Lund}, {Silva Aguirre},
  {Arentoft}, {Kjeldsen}, {Albrecht}, {Chaplin}, {Isaacson}, {Pedersen},
  {Jessen-Hansen}, {Tingley}, {Christensen-Dalsgaard}, {Aerts}, {Campante}, \&
  {Bryson}}]{vaneylen14}
{Van Eylen}, V., {Lund}, M.~N., {Silva Aguirre}, V., {et~al.} 2014, \apj, 782,
  14

\bibitem[{{Weiss} \& {Marcy}(2014)}]{weiss14}
{Weiss}, L.~M., \& {Marcy}, G.~W. 2014, \apjl, 783, L6

\bibitem[{{Winn} {et~al.}(2010){Winn}, {Fabrycky}, {Albrecht}, \&
  {Johnson}}]{winn10}
{Winn}, J.~N., {Fabrycky}, D., {Albrecht}, S., \& {Johnson}, J.~A. 2010, \apjl,
  718, L145

\bibitem[{{Xie} {et~al.}(2016){Xie}, {Dong}, {Zhu}, {Huber}, {Zheng}, {De Cat},
  {Fu}, {Liu}, {Luo}, {Wu}, {Zhang}, {Zhang}, {Zhou}, {Cao}, {Hou}, {Wang}, \&
  {Zhang}}]{xie16}
{Xie}, J.-W., {Dong}, S., {Zhu}, Z., {et~al.} 2016, Proceedings of the National
  Academy of Science, 113, 11431

\bibitem[{{Zerbi} {et~al.}(1999){Zerbi}, {Rodr{\'{\i}}guez}, {Garrido},
  {Mart{\'{\i}}n}, {Arellano Ferro}, {Sareyan}, {Krisciunas}, {Akan}, {Evren},
  {Ibano{\v g}lu}, {Keskin}, {Pekunlu}, {Tunca}, {Luedeke}, {Paparo}, {Nuspl},
  \& {Guerrero}}]{zerbi99}
{Zerbi}, F.~M., {Rodr{\'{\i}}guez}, E., {Garrido}, R., {et~al.} 1999, \mnras,
  303, 275

\end{thebibliography}

\end{document}